# Gated Mode Superconducting Nanowire Single Photon Detectors

Mohsen K. Akhlaghi[*,1] and A. Hamed Majedi[*,1,2]

[1]Institute for Quantum Computing and ECE Department, University of Waterloo, 200 University Ave West, Waterloo, ON, Canada, N2L 3G1

[2]Harvard School of Engineering and Applied Sciences, 29 Oxford Street, Cambridge, MA 02138, USA

[*]e-mail: mkeshava@maxwell.uwaterloo.ca; ahmajedi@maxwell.uwaterloo.ca

**Single Photon Detectors (SPD) are fundamental to quantum optics and quantum information. Superconducting Nanowire SPDs (SNSPD) [1] provide high performance in terms of quantum efficiency (QE), dark count rate (DCR) and timing jitter [2], but have limited maximum count rate (MCR) when operated as a free-running mode (FM) detector [3, 4]. However, high count rates are needed for many applications like quantum computing [5] and communication [6], and laser ranging [7]. Here we report the first operation of SNSPDs in a gated mode (GM) that exploits a single photon triggered latching phenomenon to detect photons. We demonstrate operation of a large active area single element GM-SNSPD at 625MHz, one order of magnitude faster than its FM counterpart. Contrary to FM-SNSPDs, the MCR in GM can be pushed to GHz range without a compromise on the active area or QE, while reducing the DCR.**

When a superconducting nanowire is biased close to its critical current ($I_C$), a photon induced resistive hotspot followed by current assisted formation of a resistive bridge, results in a macroscopic voltage pulse [1]. SNSPDs provide outstanding performance in a small cryocooler [8], and thus are the most practical cryogenic SPD. QE up to 57% at 1550nm [9], DCR below 1Hz [10] and timing jitter less that 30ps [11] have been reported. Furthermore, it is possible to resolve the number of photons by spatially multiplexing many nanowires in different configurations [11, 12]. Quantum key distribution [6], lunar laser communication link [13], diagnosis of integrated circuits [14] and characterization of single photon sources [15] are among the demonstrated applications.

To date SNSPDs have been operated in FM, i.e. the nanowires are in superconducting phase and are biased by a constant current as shown by the circuit model in Fig. 1a [16]. For proper FM operation, the system should be mono-stable in superconducting phase to ensure the resistive bridge would always annihilate after formation and the SNSPD would self-reset to its initial state. The smaller the time constant $\tau_e = L_K/R_L$, the faster the return to initial state and the higher the MCR [16]. However, current biased superconducting thin-film wires are thermally bistable systems in general [17, 18]. Keeping the QE at maximum, mono-stability condition puts a lower limit on $\tau_e$ ($\tau_{e\text{-min}}$) beyond which the FM-SNSPD would turn into a bistable system and therefore would stop working by permanently latching to a stable resistive state [3, 4]. Furthermore, $\tau_{e\text{-min}}$ scales up with $L_K$ and consequently larger single element FM-SNSPDs have less MCR.

There have been different approaches to reduce $\tau_{e\text{-min}}$, while keeping the active area unchanged. It can be reduced by exploiting different materials [19, 20, 21], or by changing the geometry of the device from a single long nanowire to some shorter wires that are either connected in parallel [22, 23, 24], operating as independent smaller pixels [11] or placed under a nano-antennae [25]. However, keeping the structure and materials unchanged, the tradeoff between the active area and MCR remains intact. In the following, we show in the GM-SNSPDs, the condition $\tau_e > \tau_{e\text{-min}}$ can be relaxed and thus the MCR would be enhanced.

In GM, we increase $R_L$ such that $\tau_e < \tau_{e\text{-min}}$ to make an electrically fast but bistable system. However, instead of a constant bias current, we apply an alternating current with a DC offset to the nanowires. At the peak of the current, the superconducting nanowire latches to the resistive state upon a photon detection or a dark count. The latched nanowire will then be reset to superconducting state by the next minima of the current. Therefore, latching to the resistive state which has been a forbidding effect for FM-SNSPDs, would be part of the detection process in GM. Also, because $\tau_e < \tau_{e\text{-min}}$, a GM-SNSPD would have higher MCR compared to the same SNSPD operated in FM.

We use the circuit shown in Fig. 1b to implement a GM-SNSPD. The voltage from a signal source is split into two paths. One path creates the alternating bias current by using a biasing resistor, $R_B$. This current is sensed by $R_2$ together with two amplifiers. The other path undergoes an adjustable delay and attenuation. The difference signal, $V_d=V_2-V_1$ would be small in absence of incoming photons for an appropriately adjusted circuit. However, as illustrated in Fig. 1c, $V_d$ jumps up whenever the detector latches. We use discriminated $V_d$ to count photons. We also make an FM-SNSPD to compare it with GM operation: a DC signal source is used in the same circuit of Fig.1b but $R_1$ replaced with a large 100nF capacitor. Such circuit is electrically equivalent to the one in Fig. 1a with $R_L=R_B+R_2$ and thus provides FM operation with minimal changes.

The devices used in this work were fabricated by Scontel, Moscow, Russia. They consist of a single 500μm long, 4nm thick, 120nm wide NbN on Sapphire. Active area is 10x10 μm$^2$ (see [26] for a picture). At 4.2K, we determined $L_K≈490$nH (see methods) and $τ_{e-min}≈3.3$ns (equivalent to $R_L=150Ω$) (see methods).

We excite the detectors using an attenuated 1310nm CW laser at a level that makes the measured count rates linearly proportional to the laser intensity, thus ensuring single photon sensitivity [27]. Both the discriminated time binned response of the FM-SNSPD, and the discriminated response of the GM-SNSPD make binary random process x, taking 0 for no-click and 1 for click events. The autocorrelation function, $Γ(τ≠0)$ of x gives the joint probability of two events separated in time by τ. We use normalized $Γ(τ≠0)$ to check the independency of successive events and therefore to study features like the dead time and after pulses.

Fig. 2a shows the measured $Γ(τ)$. For GM: $R_B = 650Ω$, the bias is 625MHz sinusoidal with minima and maxima equal to -2μA and $0.9I_C$ respectively (see methods). For FM: allowing a margin to further avoid latching, we set $R_L = 100Ω$, the DC bias is $0.9I_C$. The results show for having $Γ(τ)$ changed by less than ±10%, τ should be greater than 22ns in FM, while within the same limits, adjacent gates of 625MHz GM-SNSPD keep their statistical independency. This shows about one order of magnitude speed up in GM.

Also, Fig. 2b is a time histogram of the detection events within a gate period. It shows the QE changes less than 5% for a time window equal to 57ps (about 1/30th of the gating period).

To study the after pulses in GM, we keep the biasing unchanged but excite the detector with a 625MHz/20, 1310nm pulsed laser. The resulting normalized $\Gamma(\tau)$ is shown in Fig. 2c. It shows clear jumps each 20 gating periods and, in between, remains flat at a level determined by the dark count probability per gate. An exception occurs at the first gate where it is enhanced by an after pulsing probability of about 0.03%. We attribute this to either an unwanted oscillatory behavior in the biasing current following a photon detection or temperature rise in the corresponding gate. The possibility of both options will be seen later in the paper.

QE and DCR of both of the modes are compared in Fig. 3. For GM we apply 100MHz bias and lock a 200ps, 1310nm pulsed laser to the bias maxima. QE for GM and FM shows good agreement. However, as the GM-SNSPD is not always on, its DCR is smaller than that of FM-SNSPD by about one order of magnitude. As we couldn't synchronize our laser to the bias current oscillation at higher frequencies, we excite the detector with a CW laser. In this way, we confirmed both the average QE (i.e. detection probability per gate divided by the average number of photons per gate period) and the DCR stayed unchanged when the bias frequency was shifted to 625MHz.

To explore the limitations on the gating frequency, we developed an electrical model of our SNSPD as shown in Fig. 4a (see methods). An approximate version (see Fig. 4b), plus the thermal model in [28] is used to numerically simulate our GM-SNSPD. We simulate the peaks of current in the gates following photon detection. The result is shown in Fig. 4c for $R_p$ = 725Ω (equivalent to $R_B$=650Ω). At less than about 300MHz or at about 600MHz the peaks do not change significantly. Indeed, this is how the gating frequencies in the experiments are selected. Therefore, the oscillatory response of an under-damped RLC circuit puts a purely electrical limitation for the gating frequency of our GM-SNSPD.

To study the ultimate MCR of GM-SNSPDs, assuming $C_P$ can be reduced to 0.01pf if $R_B$ be integrated to the SNSPD chip, we repeat the simulation for values of $L_K$ ranging 6nH to 6µH. For each $L_K$, we choose $R_B$ such that it makes a critically-damped RLC circuit. We simulate for the maximum temperature of the nanowire at the center of the gate following a detection gate. The results are shown in Fig. 4c. The curves are up to the frequency at which the detector re-latches in the next maxima of the current due to elevated nanowire temperature. We also confirm that the currents of Fig. 4.c are horizontally flat for this case. Therefore, for such an integrated GM-SNSPD the MCR would be purely limited by the thermal response of the SNSPD. Notice, increasing $L_K$ over three orders of magnitude decreases the MCR for about 33%. Therefore, in contrast with FM-SNSPDs, no compromise on active area is required for achieving high MCR in GM.

To conclude, SNSPDs can be operated in GM at the same QE that they have in FM, but with an enhanced MCR and reduced DCR. Using a differencing read out technique, we implemented a gated setup and characterized different features of it. We have shown how irrespective of the value of $L_K$, the MCR can be pushed to GHz range where a purely thermal limitation does not let faster operation. The work will add a degree of freedom for designing ultra-high speed SPDs for applications like quantum key distribution and laser ranging.

**Methods**

**SNSPD Electrical Model –** We used high frequency electromagnetic simulation software, SONNET, to derive the SNSPD circuit model as shown in Fig. 4a. Perfect conductor on top of a loss-less substrate with relative permittivity equal to 11.35 [29, 30] was assumed to simulate for the Gold pads. Surface inductance equal to 90pH (equivalent to London Penetration depth equal to 532nm [16] at thickness of 4nm) was used to simulate the meandering nanowires. In each case we simulated for the phase and amplitude of the S-parameters up to 5GHz. These results are then converted to the circuit model by using Agilent Advanced Design System (ADS). The 0.14pf capacitor is obtained with the same methods for a small pad on a printed circuit board where the SNSPD was connected to.

To test the validity of the resulting model, we put $R_B = 50\Omega$, disconnect the connection to the HEMT amplifier and connect $RF_{in}$ to a coaxial cable. We measured the input reflection coefficient using a vector network analyzer (VNA) calibrated at the cryogenic end of the coax while the device was cooled to 4.2K. The agreement between the measurement and what we simulate in ADS using the model in Fig. 4a, confirmed the model is fairly accurate at least up to 2GHz.

**Minimum $\tau_e$ in free running mode –** We adapted the same method reported in [3] to measure the maximum available count rate of the FM-SNSPD while maintaining the QE at its maximum. For different $R_L$ values, starting from a high bias current that results in a stable latched state, we measured the current at which the nanowire returns to superconducting state while sweeping the bias current downwards. We observed the return current starts to decrease for $R_L$ greater than about 150Ω. This value and its associated $\tau_e$ are what we report as $\tau_{e-min}$ of our FM-SNSPD.

**Adjusting the high frequency current –** One of the difficulties of operation in GM is accurately applying a high frequency biasing current with a DC offset to a device operating at cryogenic temperature. Using a VNA, we characterized the components of our electrical setup including biasing coaxial cable and all individual electronic components used. The resulting information was all put together in Agilent ADS where we simulated for frequency dependence of the Transconductance between the voltage generated by the signal source and the current that flows in the Nanowires. This was used to adjust the minima and maxima of the biasing current throughout the experiments.

It is better to set the minima of current to zero to ensure the Nanowire always returns to superconducting state following by a latching. We chose a slightly negative value (-2µA) for the minima to allow more room for the effect of noises and also errors in determining the Transconductance.

**Acknowledgements**

We acknowledge the financial support of OCE, NSERC and IQC. The authors would like to acknowledge Jeff. S. Lundeen and Thomas Jennewein for helpful comments.

**Author Contributions**

A.H.M. inspired, supervised, provided the funding and contributed to conclusions, discussions and writing the paper. M.K.A. designed and performed the experiments, did the modeling and simulations and wrote the paper.

**Competing Financial Interests Statement**

The authors declare no competing financial interests.

**Figure Legends**

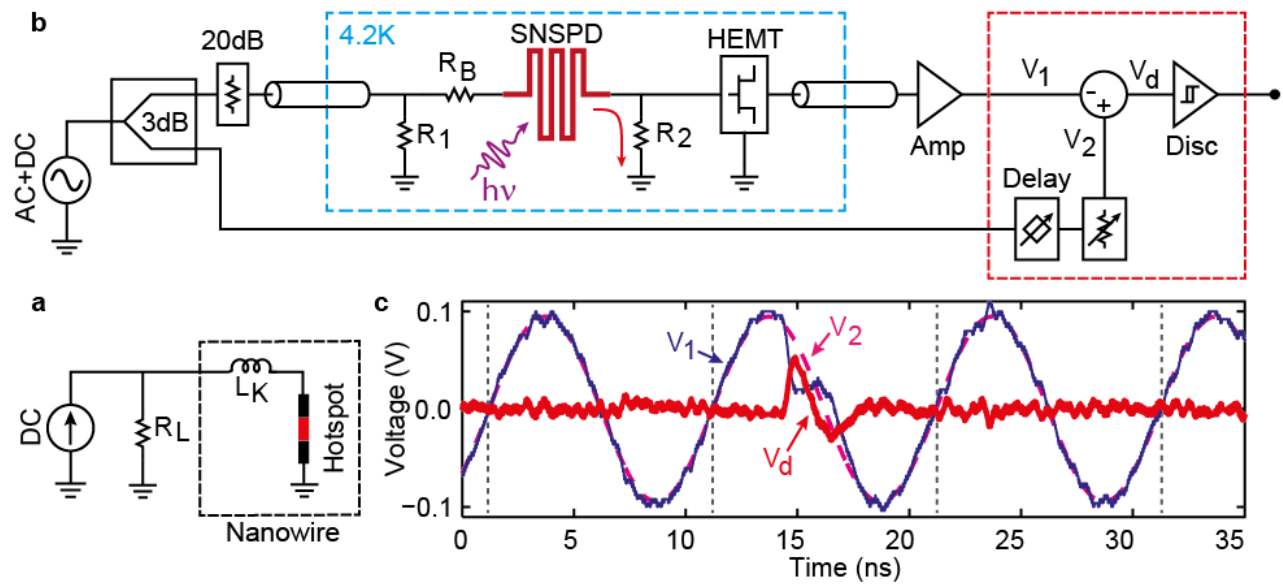

**Figure 1 Gated mode operation of SNSPDs. a,** The equivalent circuit model for a FM-SNSPD. $L_K$ is the nanowire kinetic inductance and $R_L$ is the load seen by the nanowire. **b,** Schematic for our GM-SNSPD showing the major elements. 3dB labels a 3dB power splitter and 20dB labels a 20dB resistive attenuator. $R_1$ is a 50Ω load resistor to terminate the coax and $R_2$ is a 50Ω current sense resistor. We used a high electron mobility transistor (HEMT) to both amplify the weak signal and to further isolate the Nanowire from the reflected signals. **c,** A typical set of waveforms showing the detector latches at the current maxima, returns to the superconducting state at a smaller current and the difference signal peaks as a result.

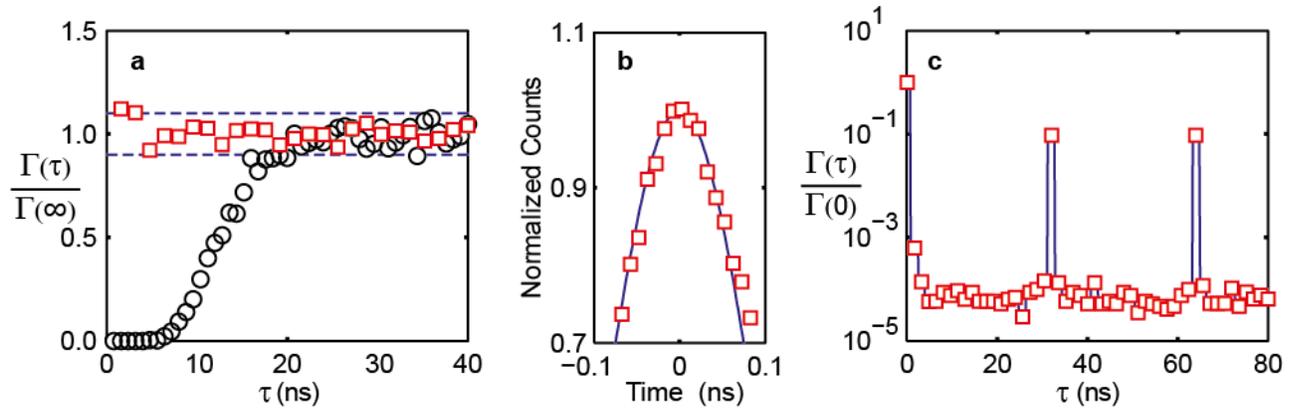

**Figure 2 Autocorrelation and gate shape measurements. a,** Normalized Γ(τ) for FM-SNSPD with $R_L$ = 100Ω (black circles) and our 625MHz GM-SNSPD (red squares), both under CW laser illumination. Non-flat autocorrelations show dependency of two detector clicks separated in time by τ. **b,** Normalized time histogram of detection events within a gate period of our 625MHz GM-SNSPD under a CW laser. **c,** Normalized autocorrelation for 625MHz GM-SNSPD excited with a pulsed laser at 625MHz/20.

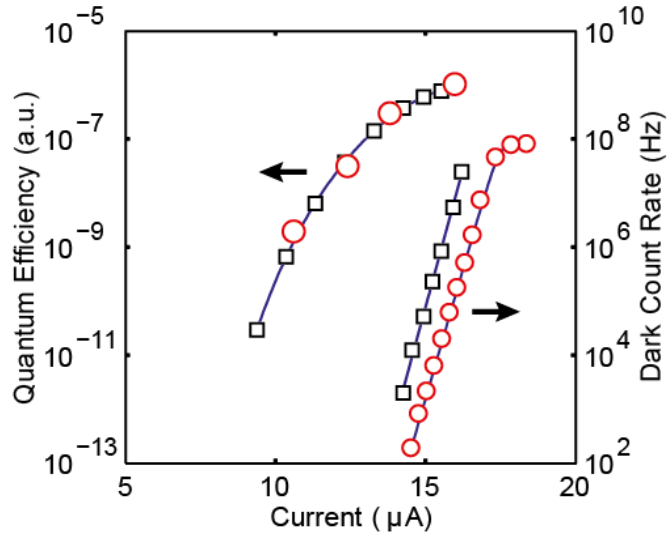

**Figure 3 Quantum Efficiency and Dark Count measurements.** Black squares are for the FM-SNSPD and red circles are for the 100MHz GM-SNSPD. The current shown is the DC biasing current or the peak of current for FM and GM respectively. For GM QE measurement, a 200ps pulsed laser was locked to the peaks of the current through the SNSPD. Note for GM-SNSPD the DCR saturates at the gating frequency at higher currents.

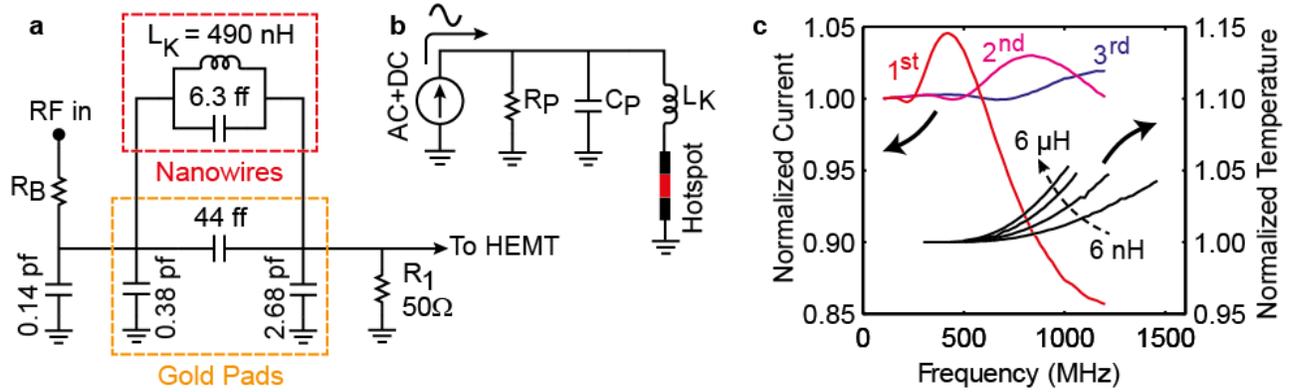

**Figure 4 Electro-thermal model and simulation. a,** The electrical circuit model for our GM-SNSPD (see methods). **b,** A simplified version that was used to do electro-thermal simulations. **c,** Simulated peaks of current in the 1st, 2nd and 3rd gate following a photon detection normalized to 95% of the critical current for $R_P$=725Ω (equivalent to $R_B$=650Ω), $C_P$=0.57pf and $L_K$=490nH. Also shown is the maximum temperature on the surface of the Nanowire at the first gate following a photo-detection normalized to 4.2K for a critically damped circuit with $C_P$=0.01pf and $L_K$ equal to 6nH, 60nH, 600nH and 6000nH.